\documentclass[usenatbib]{mn2e}
\usepackage{amsfonts,amssymb,amsmath}
\usepackage{graphicx}
\usepackage{url}
\title[Hydrodynamical model of A0620-00 light curve]
{Hydrodynamics associated to the X-ray light curve of A0620-00}

\author[Y. Coronado \& S. Mandoza]{Y. Coronado,
S. Mendoza\thanks{E-mail~address:~coronado@astro.unam.mx,~sergio@astro.unam.mx.} \\
Instituto de Astronom\'ia, Universidad Nacional Aut\'onoma de
M\'exico, AP 70-264, Distrito Federal 04510, M\'exico}

\begin{document}


\pagerange{\pageref{firstpage}--\pageref{lastpage}} \pubyear{2013}

\maketitle

\label{firstpage}

\begin{abstract}
  From 1975 to 1976, an outburst was detected in the light curve of
the X-ray transient A0620-00 using the Ariel~V and SAS-3 experiments.
In this letter we model the outburst with the hydrodynamical model
proposed by Mendoza et al. (2009).  The physical model is constructed
assumming basic mass and momentum conservation laws  associated to the
motion of the shock waves developed inside the expanding relativistic
jet of the source.  These internal shock waves are produced as a result
of periodic variations of the inyected mass and velocity of the flow at
the base of the jet. The observations of this  X-ray light curve present
two clear bumps.  The first one is modelled assuming periodic variations
of the inyected velocity at the base of the jet, while the second one can
either be modelled by a further velocity oscillations, or by a periodic variation
of the mass injection rate at the base of the jet at a latter time.
The fitting of the data fixes different parameters of the model, such as
the mean mass injection rate at the base of the jet and the oscillation
frequency of the flow as measured on the rest frame of the central source.
\end{abstract}


\section{Introduction}
\label{introduction}

  On August 3rd, 1975 the low-mass X-ray binary black hole transient
A0620-00, exhibited its most powerful outburst detected by the Sky Survey
Experiment on board the Ariel~V satellite in X-rays \citep{Elvis75}. On
August 8th, this micro-quasar was also followed by the SAS-3 X-ray
observatory \citep{Matilsky76}.  
Subsequently it was also seen in different wavelengths,
from radio to ultraviolet \citep[see][for a review]{kuulkers98}.  At that
time, A0620-00 became the most powerful X-ray source in the sky for
almost two months.

  Five days after the discovery of A0620-00 intense variations on time
scales of days, which reached a maximum value \(\sim 50\) times that
of the Crab Nebula in the energy interval of \( 1.5-6 \)~KeV, suggested
that the source was an an excellent candidate for a stellar mass black
hole with a stellar companion.  This idea was further corroborated by
the direct observations made by \citet{McClintock86} which resolved the
binary components of the source.  The estimated distance to A0620-00 is
\( \sim \! 1\textrm{Kpc}\) \citep{Shahbaz94}, being one of the nearest
X-ray transients objects, hosting a black hole with a mass function
\( f(m) = 3.18 \pm 0.16 \, M_{\odot}\) \citep{McClintock86,marsh94}.

Using dynamical and stellar numerical models, the inclination of the
accretion disc with respect to the orbit spanned by the black hole and the
stellar companion yields \( i = 51^{\circ} \pm 0.9\), implying
a black hole mass \( 6.6\pm 0.25 \, M_{\odot} \), and an  estimated
distance to the source \( 1.06 \pm 0.12 \, \textrm{Kpc}\)~\citep{Cantrell10}.

 The radio emission of A0620-00 was detected in 1975 \citep{Davis75,
Owen76},  with no jet resolved.  Since many X-ray transient systems
containing a black hole have radio emission that follows their X-ray
outburst with clear detections of relativistic outflows or jets
\citep{2009Sci...323.1688A}, it was clear that a jet should have been
produced in the X-ray outburst of A0620-00. \citet{Kuulkers99} infered
the existance of that jet by compiling different radio observations,
concluding that the speed of the jet \( \sim 0.9 c \), where \( c \)
is the velocity of light.

In this letter, we assume that the mechanism producing the observed
light curve of A0620-00 is caused by variations in the injected flow
at the base of the jet, which leads to the formation of shock waves
that propagate along the jet. The hydrodynamical jet model presented
in \citet[][hereafter M09]{mendoza09} describes the motion of working
surfaces inside a relativistic jet, which are able to fit the observed
light curves of long gamma-rays bursts (lGRBs) as  well as the light
curve of the blazar PKS~1510-089 \citep{cabrera13}. The shape of the X-ray
light curve of the micro-quasar A0620-00 is similar to the one observed in
lGRBs, showing an exponential rapid increase with a slow decay.  With all
these, the the physical ingredients of the phenomena that produces the
light curve of A0620-00 can be considered similar to those ones ocurring
in lGRB and on PKS~1510-089, but with different physical scales of energy,
sizes, masses, accretion power rates, etc. \citep{mirabel02}.

The letter is organised as follows. In Section~\ref{data} we present
the X-ray data of the light curve of A0620-00. In Section~\ref{model}
we briefly describe the main features of the hydrodynamic model by M09, and
using that model we fit the observational data in Section~\ref{sec:Fit}.
Finally, the results of our fits and the discussion of the main
physical parameters inferred by the modelling are presented in
Section~\ref{discussion}.

\section{Observational data}
\label{data}

\begin{figure}
\begin{center}
 \includegraphics[scale=0.65]{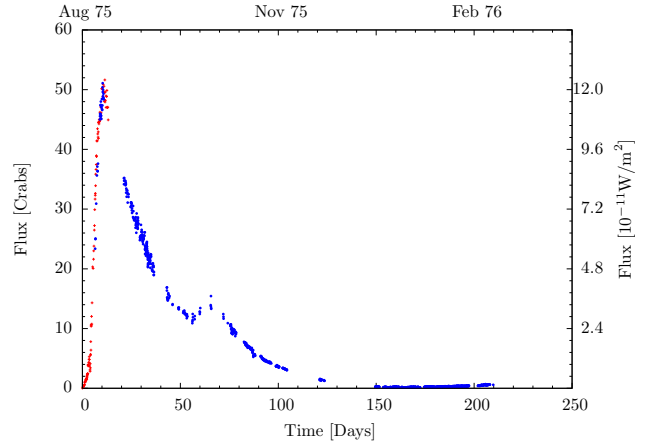}
\end{center}  
\caption{The figure shows the X-ray light curve of the micro-quasar
A0620-00.  The crossess correspond to the Ariel~V observations,
which covers the uprise of the curve and the beginning of its decay.
The points are SAS-3 observations, which cover the first outburst
and the bump at the decaying of the burst.
}
\label{fig1}
\end{figure}

The observational 1975-1976 X-ray light curve of the micro-quasar
A0620-00 is shown in Figure~\ref{fig1} and was kindly provided by
Jeffrey McClintock.  It consist of a composition of two independent
lightcurves obtained by \citet{Elvis75} and \citet{Matilsky76},
with instruments on board Ariel~V and SAS-3 respectively.  Both data
count-rates have been converted to flux Crab units, according to the
instruments specifications \citep{Whitlock92}.  With this it is possible to get a complete
light curve of the 1975-1976 outburst, including a bump in the decaying
outburst. Figure~\ref{fig1} shows the plotted data on a linear scale, with
the advantage of revealing the impressive outburst of 1975 and a clear
second bump a few days after the maximum.  To convert from Crabs to~mJy,
we use the conversion given by \citet{kirsch05} and \citet{Bradt79}
for the Ariel~V data (in the energy range \( 1 - 13 \)~KeV) and the one
in \url{http://heasarc.gsfc.nasa.gov/docs/sas3} for the SAS-3 satellite
(in the energy range \( 2 - 10 \)KeV).  This conversion is coherent with
the results obtained by \citet{kirsch05}, for which \( 1 \text{Crab}
\approx 2.4 \times 10^{-11} \mathrm{W} \mathrm{m}^{-2} \) in the energy
range \( 1 - 13 \mathrm{KeV}\).

  In order to calculate the Luminosity \(L\) we multiply
the obtained Flux \(F\) by \( 4\pi D^2 \delta^{-p}\),
where \(\delta:=1 / \Gamma(v_0)(1-(v_0/c)\cos\theta)\). 
For this particular case, since the inclination
angle \( i \approx 51^\circ \), then the angle \( \theta \)
between the jet and our line of sight is \( \approx 39^\circ\), with a 
distance to the source  \( D = 1 \,\textrm{Kpc}\)
\citep{1976ApJ...203L..15O,Shahbaz94,Cantrell10}.
The beaming index \( p \) for synchrotron radiation is \( 3 \)
\citep{longair} and we have chosen such value in accordance to the
calculations of blazars and lGRBs \citep{wu11,mendoza09,cabrera13}, having
in mind a unified radiative model for the flow inside all relativistic
astrophysical jets. With this luminosity, and with  the average jet
bulk speed of \( v_{0}=0.9
c \) \citep{Kuulkers99}, we are able to fit the observational
data with the hydrodynamical model of M09.

  Attempts to model the light curve of A0620-00 were made by
\citet{Kuulkers99} who noticed that this behaviour might well be
understood modelling many ``\emph{synchrotron bubble}'' ejections.
Since micro-quasars are thought to be short scaled versions of
quasars  and are thus logical scaled counterparts of
lGRB~\citep{mirabel02}, it is quite natural to model their behaviour
using the model by M09 to model their light curve.  We thus assume that
velocity and mass variations at the base of the jet of the micro-quasar
A0620-00 produce internal shock waves that travel inside the expanding
relativistic jet and that these shock waves in turn are able to reproduce
its observed light curve.

%
\section{The hydrodynamical model. }
\label{model}

 Many relativistic jets show internal shock waves, which are due to the
interaction of the jet with inhomogeneities of the surrounding medium
\citep[see e.g][]{mendoza-1-01},
 the bending of jets \citep[see e.g][]{mendoza-1-02} and time
fluctuations in the velocity and mass of the ejected material
\citep[cf.][]{rees94,jamil08,mendoza09}.  In particular the
semi-analytical model of M09 is a hydrodynamical description of time
fluctuations at the base of the jet that develop shock waves inside an
expanding relativistic jet.

  The model of M09 produce internal shock waves by periodic oscillations
of speed and mass discharge at the base of the jet.  This mechanism
injects fast fluid that overtakes slow one, producing an initial
discontinuity which eventually forms a working surface expanding along
the jet.  The extra kinetic energy inside the working surface is thus
radiated away.  The efficiency converting factor between kinetic energy
and observed radiation is assummed to be \( \sim 1 \).  This value was
used by M09 and \cite{cabrera13} for lGRBs and the Blazar PKS1510-089. We
have made such a choice, since a micro-quasar can be considered as
a scaled version of a quasar.  Furthermore, A0620-00 has been the
most energetic X-ray micro-quasar and in this sense shares the same
behaviour as PKS1510-089 which presented extreme \(\gamma\)-ray energy
detections. As explained in Section~\ref{introduction},
the micro-quasar A0620-00 behaves as an scaled typical lGRB and as such,
the hypothesis used by M09 can be extended to this particular object.
As we will discuss in section~\ref{discussion}, this assumption yields
physical parameters which are coherent with the expectations of typical
micro-quasars.

  Following M09, we assume that the flow is injected at the base
of the jet with a periodic velocity \( v \) given by:

\begin{equation} 
  v(\tau)= v_0 + c \eta^2 \sin \omega \tau, 
\label{eq01}
\end{equation}

\noindent and a periodic mass injection rate:

\begin{equation} 
  \dot{m}(\tau)= \dot{m}_0 + \epsilon^2 \sin \Omega \tau, 
\label{eq02}
\end{equation}

\noindent where \( \tau \) is the time measured in the proper frame
of the source, the velocity \( v_0 \) is the ``average'' velocity of
the flow inside the jet, and \( \omega \) is the oscillation frequency
of the flow. The positive constant parameters \( \eta^2 \)  and \(
\epsilon^2 \) are obtained by fitting the observational data, with the
particular feature that  \( \eta^2 \) has to be sufficiently small so that
the bulk velocity \( v(\tau) \) does not exceed the velocity of light
\( c \).  The mass injection rate \( \dot{m}_0 \) is the ``average''
discharge of the flow at the base of the jet, and \( \Omega \) is its
oscillation frequency.

\section{Modelling the X-ray light curve} 
\label{sec:Fit}

  As previously discussed, the first outburst resembles
the light curve of a typical lGRB. As such, we model that burst by
assuming \( \dot{m} = \textrm{const.} \), in complete accordance to
the calculations by M09. The bump in the decay of
the first burst is modelled in two ways.  The first is by assuming a
new ejection with constant discharge added up to the first outburst.
The second way is by assuming an oscillating mass discharge \( \dot{m}
\) produced at a particular time while the first outburst decays.

  In the first burst, where \( \dot{m} = \textrm{const.} \), the 
semi-analytical model presented by M09, requires to know the values
of \( v_0 \), \(\eta^2 \),  \( \omega \) and \( \dot{m} \).
The ``mean'' velocity value \(v_0\) can be taken from observational data.
For this particular case,  we choose the inferred value from a wide variety
of radio observations modelled through ejection mechanisms 
by \citet{Kuulkers99} which yields a Lorentz factor $\Gamma(v_0)=2.3$.
Since the value of \(\eta^2 \) has to be small due to the expected variations
inside the jet, we start with a small value of \(\eta^2\) such that
the bulk velocity of the flow \( v(\tau = 0)  \sim 0.1 \times v_{0}\).
The velocity variations \( v(\tau) \) are thus allowed to vary from
this value up to the extreme upper limit \( \Gamma(v(\tau)) \sim 10 \).
As pointed by M09, the mass ejection rate is related to the observed
luminosity \(L = \dot{m}c^{2}\) and is obtained directly from the fits
of the light curve.

  The second burst can be described by two different mechanisms: (a)
The mass discharge \( \dot{m} \) is kept constant and the velocity is the
sum of the velocity as in equation~\eqref{eq01} with an extra oscillating
term \( \eta'^2 \sin \omega' \tau \), with the same \(v_0 \), \( \eta^2
\) and \( \omega \) used for the callibration of the first outburst.
(b) The velocity is the same as the one used for the callibration of
the first burst, and the mass discharge \( \dot{m} \) is allowed to
oscillate as in equation~\eqref{eq02}, with \( \dot{m}_0 \) given by
the results obtained with the callibration of the first outburst.

  Following \citet{cabrera13}, we set a dimensionless system of units to
perform the required fitting.  To do so, the luminosity is measured in
units of the peak luminosity and the time in units of the FWHM of each
particular outburst.  This system of units is such that for the first
outburst \( \omega = 1  \) and \( \dot{m} = 1 \), with the only unknown \(
\eta^2 \) obtained by a linear regression analysis to within \(10 \)\%
of accuracy.  For the case of the second outburst: (a) The only unknown
is \( \eta'^2 \) obtained with a further linear regression analysis. (b)
The unknown quantity is \( \epsilon^2 \) which can be obtained by another
regression analysis.  To return to the physical system of units one can
recall at any particular step that the luminosity \( L = \dot{m} c^2 \)
(for the first outburst) and that the time \( t = \omega^{-1} \tau \)
(case (a) of the second outburst),  \( t = \omega'^{-1} \tau \) and \(
t = \Omega^{-1} \tau \) (case (b) of the second outburst).

\begin{table}

 \begin{center}
 \begin{footnotesize}
 
 \begin{flushleft}
 \fbox{\textbf{1st. outburst}}
 \end{flushleft}
 \begin{tabular}[0.75\textwidth]{p{2cm}p{2cm}p{2cm}c|c|c|c }
\hline
$ \eta^2 / c $   & $ \omega $  &  $ \dot{m} $ & $ \Gamma_\text{max} $ \\
$ 10^{-3} $	 & $ 10^{-2}$d &  $10^{-9} M_\odot \text{yr}^{-1} $ &  \\
\hline
$ 1.679 $        & $ 6.6 $     & $ 2.8063 $   & $ 2.31 $              \\
\hline	
 \end{tabular}
\begin{flushleft}
  \fbox{\textbf{2nd. outburst - case (a)}}
\end{flushleft} 
 \begin{tabular}{p{2cm}p{2cm}p{2cm}cccc}
 
\hline
$ \eta'^2 / c $   & $ \omega' $  &  $ \dot{m} $ & $ \Gamma_\text{max} $ \\
$ 10^{-3} $	 & $ 10^{-2}$d &  $10^{-9} M_\odot \text{yr}^{-1} $ &  \\
\hline
$ 0.061 $        & $ 249.1 $     & $ 0.8391 $   & $ 3.61 $              \\
\hline
 \end{tabular}
 \begin{flushleft}
  \fbox{\textbf{2nd. outburst - case (b)}} 
 \end{flushleft}
  \begin{tabular}{p{2cm}p{2cm}p{2cm}lccr}
\hline
$ \epsilon^2 / c $   & $ \Omega $  &  $ \dot{m} $ & $ \Gamma_\text{max} $ \\
$  10^{-9} M_\odot \text{y}^{-1} $ & $ 10^{-2}$d  &  $10^{-9} M_\odot \text{yr}^{-1} $ &  \\
\hline
$ 0.8959  $          & $ 1.5 $     & $ 0.7466 $   & $ 2.31 $             \\ 
\hline                   
 \end{tabular}
  \caption{Obtained values for the free parameters of the model by M09
after fitting with X-ray observations of the light curve of the
micro-quasar A062-00, accurate to within \(10\)\%.  The background Lorentz 
factor of the bulk velocity of the flow was assumed to be \( 2.29 \). 
The maximum Lorentz factor of the flow in each outburst is represented by
\( \Gamma_\text{max} \), and 
the minimum is \( \sim 1.8 - 2.2 \).
}
 \label{table1}
 \end{footnotesize} 
 \end{center}
\end{table}

\begin{figure}
\begin{center}
 \includegraphics[scale=0.65]{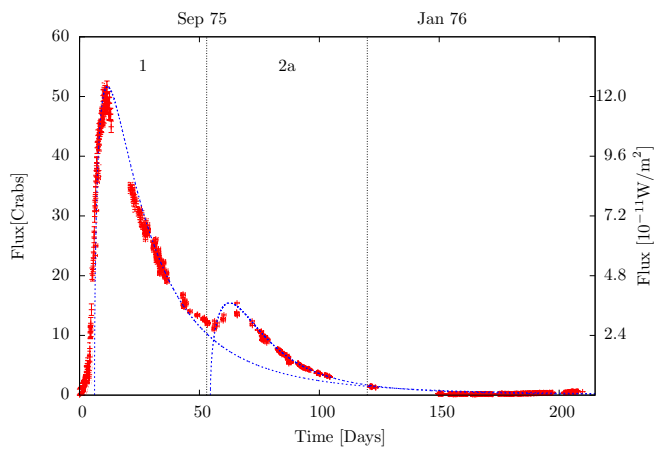}
\end{center}  
\caption{The figure shows the fits to the X-ray light curve observations of
the micro-quasar A0620-00, which
corresponds to velocity variations and constant mass discharges for the 
first and second outburst (model (a) -see text).  The second outburst 
has an additional oscillating velocity component as compared to the first
one.
}
\label{fig2}
\end{figure}

\begin{figure}
\begin{center}
 \includegraphics[scale=0.65]{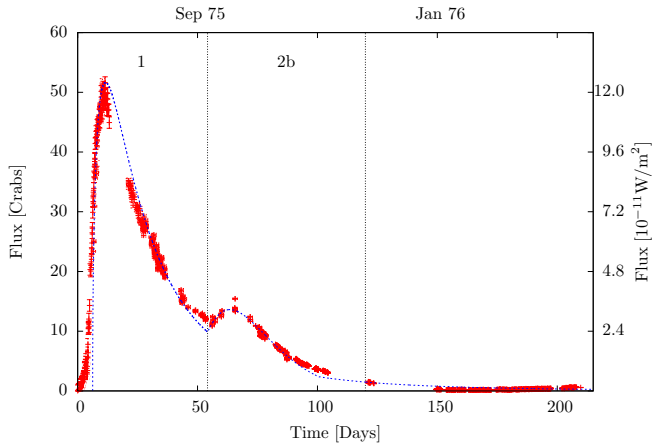}
\end{center}  
\caption{ The figure shows the fits to the X-ray light curve observations of
the micro-quasar A0620-00, which
corresponds to velocity variations for the 
first and second outburst, but with constant mass discharge at the first
outburst and oscillating mass discharge at the second outburst (model (b) 
-see text).
}
\label{fig3}
\end{figure}

\section{Discussion}
\label{discussion}

  The results of the fits to the X-ray data presented in Section~\ref{data}
using the model by M09 are shown in Figures~\ref{fig2} and~\ref{fig3}.
The obtained values for the physical parameters of the model
are presented in Table~\ref{table1}.
We have also included the maximum and minimum Lorentz factors,
obtained for the bulk velocity of the flow. 
Direct inspection on the results of the Table show that 
\( \dot{m} \sim 10^{-9} -10^{-10} \textrm{M}_\odot  \textrm{yr}^{-1} \), \(
\omega^{-1} \sim 0.01 - 2 \, \text{days} \)  with a maximum Lorentz
factor \( 2.3 - 3.6 \).

  A00620-00 resulted to be an ideal target to test the model by
\citet{mendoza09} since it closely resembles a lGRB in this outstanding
outburst in x-rays. Future tests of the model have to be done with a
wide variety of Light Curves from a large collection of micro-quasars.

\section{ACKNOWLEDGMENTS}
The authors gratefully acknowledge the kindness of Jeffrey E. McClintock
for providing the observational data and for pointing to the relevant
articles discussing those observations.  This work was supported a
DGAPA-UNAM grant (PAPIIT IN111513-3).  YUC and SM thank support granted
by CONACyT: 210965 and 26344.

%
%
\bibliographystyle{mn2e-extra}
\bibliography{letterA}
\label{lastpage}

\end{document}